\documentclass[reprint,twocolumn,superscriptaddress,showpacs,nofootinbib,notitlepage]{revtex4-1}

\usepackage{graphicx}
\usepackage{latexsym,amsmath,amssymb,lmodern,float,url}
\usepackage{natbib}
\usepackage{color}
\usepackage{microtype}
\usepackage{slashed}
\usepackage{multirow}
\usepackage{bbold}
\usepackage{comment}

\newcommand{\eq}[1]{Eq.~(\ref{eq:#1})}
\newcommand{\fig}[1]{Fig.~\ref{fig:#1}}

\newcommand{\mt}{{\langle m_x(t)\rangle}}

\usepackage[colorlinks=true,backref=false, linktocpage=true,
citecolor=blue,urlcolor=blue,linkcolor=blue,pdfpagemode=UseOutlines]{hyperref}

\hypersetup{%
  bookmarksnumbered=true,
  pdftitle = {},
  pdfsubject = {},
  pdfauthor = {},
  pdfkeywords = {}
}

\def\Tr {\operatorname{{Tr}}}

\renewcommand{\d}{\mathrm d}

\newcommand{\eroq}{E$\rho$OQ}

\begin{document}
\title{Simulation of Nonequilibrium Dynamics on a Quantum Computer}
\author{Henry Lamm}
\email{hlamm@umd.edu}
\affiliation{Department of Physics, University of Maryland, College Park, MD 20742, USA}
\author{Scott Lawrence}
\email{srl@umd.edu}
\affiliation{Department of Physics, University of Maryland, College Park, MD 20742, USA}
\date{\today}

\begin{abstract}
We present a hybrid quantum-classical algorithm for the time evolution of out-of-equilibrium thermal states.  The method depends upon classically computing a sparse approximation to the density matrix, and then time-evolving each matrix element via the quantum computer.  For this exploratory study, we investigate the time-dependent Heisenberg model with five spins on the Rigetti Forest quantum virtual machine and a one spin system on the Rigetti 8Q-Agave quantum processor.
\end{abstract}

\maketitle

\section{Introduction}
Whether at the microscopic or the cosmological scale, a major challenge in physics is understanding the real-time evolution of nonequilibrium quantum systems.  Classic examples of our limited knowledge in this area are hadronization of the quark-gluon plasma produced in heavy-ion collision and the expansion of the early universe.  While in principle these problems are amenable to numerical approaches upon classical computers, the exponentially large state space of quantum systems coupled with the numerical sign problem in both fermionic systems~\cite{Troyer:2004ge} and real-time~\cite{Alexandru:2016gsd} render such calculations intractable.

The promise of quantum computers is that the computational complexity of such problems can be reduced from exponential to polynomial.  This potential improvement is two-fold: one can represent the entanglement of quantum states directly and sign-problem free real-time calculations are possible. At present, we are restricted to fewer than 50 non-error-corrected qubits, which greatly restricts the class of problems we can attempt to simulate.  Despite these present limitations, calculations in systems of interest in nuclear physics~\cite{Dumitrescu:2018njn,Roggero:2018hrn}, quantum field theory~\cite{Klco:2018kyo}, condensed matter~\cite{Macridin:2018gdw}, and quantum chemistry~\cite{lanyon2010towards,PhysRevX.8.011021} have been achieved with as few as two qubits.  Typically, these calculations have relied upon hybrid algorithms that couple a few-qubit quantum computer solving a problem of exponentially bad classical computational complexity problem to a larger classical computer.

In this paradigm, we present in this paper the Evolving Density Matrices On Qubits (E$\rho$OQ) algorithm, a hybrid quantum-classical technique for computing nonequilibrium dynamics of many-body quantum systems. In particular, we show how to compute the density matrix of a Hamiltonian $H_0$, with inverse temperature $\beta$, and then evolve this mixed state in real-time by a different (potentially time-dependent) Hamiltonian $H_1$. The algorithm proceeds by computing on a \emph{classical} computer a stochastic approximation to the density matrix $\rho = e^{-\beta H_0}$, via Density Matrix Quantum Monte Carlo~\cite{PhysRevB.89.245124}. This approximate density matrix is passed to a quantum computer element-by-element, which performs time-evolution with a different Hamiltonian $H_1$, and then computes observables with the time-evolved density matrix $\rho(t) = e^{-i H_1 t} \rho e^{i H_1 t}$.

Past theoretical work on computing thermal physics with a quantum computer has focused on performing the  thermal-state preparation on the quantum processor~\cite{2016arXiv160907877B,2010PhRvL.105q0405B}. \eroq\ differs from these approaches in allowing the computation of the thermal state to remain on the classical computer, using the quantum processor only for the classically intractable time-evolution.

In this work, we implement our algorithm for the 1D Heisenberg chain for $N\leq5$.  The real-time evolution of this system has a long history of study on classical computers, starting with~\cite{PhysRev.177.889}.  Since then, it has been used as a benchmark for developing time-dependent methods in quantum systems~\cite{PhysRevE.71.036102,PhysRevB.77.064426,PhysRevLett.93.076401,1742-5468-2004-04-P04005}.

In Sec.~\ref{sec:algorithm}, we describe the hybrid quantum-classical algorithm \eroq\ in full detail.  Following this, a brief review of the 1D Heisenberg model is covered in Sec.~\ref{sec:model}. Results using the Rigetti Forest, a quantum virtual machine (QVM)~\cite{smith2016practical}, and Rigetti's 8-qubit quantum processor (QPU) 8Q-Agave, are presented in Sec.~\ref{sec:results}, and conclusions are summarized in Sec.~\ref{sec:discussion}.

\section{The Algorithm}\label{sec:algorithm}

The first step of \eroq\ produces a stochastic, sparse approximation to the density matrix using the Density Matrix Quantum Monte Carlo algorithm (DMQMC)~\cite{PhysRevB.89.245124}, which we briefly summarize here. DMQMC is closely related to Diffusion Monte Carlo methods~\cite{anderson1976quantum}, in which a population of `psips' explore the configuration space of a system through random walks in imaginary time $\beta = i t$. Each psip is associated to a position basis state, and in the limit of large $\beta$, the density of psips approximates the ground state wavefunction. In DMQMC, the psips explore the space of basis operators, and after evolution by a finite $\beta$, the density of psips approximates the density matrix at inverse temperature $\beta$.

The density matrix $\rho(\beta) = e^{-\beta H}$ may be defined as the solution to the first-order differential equation
\begin{equation}\label{eq:bloch}
	\frac{\d \rho}{\d \beta} = - \frac 1 2 \left(H + H^\dagger\right) \rho\text,
\end{equation}
with the initial condition $\rho(0) = 1$. DMQMC stochastically implements the first-order Euler difference approximation to \eq{bloch}, with the density matrix represented by the collection of psips. To each psip is associated a basis operator $\left|b_p\right>\left<a_p\right|$ and a sign $\chi_p$, determining the sign of the psip's contribution to the density matrix.  The approximate density matrix $\tilde\rho\approx\rho$ is given by a sum over all psips: the contribution to the density matrix of each psip $p$ is $\chi_p \left|b_p\right>\left<a_p\right|$. Thus, $\tilde\rho$ is given by
\begin{equation}\label{eq:rho}
	\tilde\rho = \sum_p \chi_p \left|b_p\right>\left<a_p\right|\text.
\end{equation}
The algorithm begins by randomly placing psips along the diagonal of the density matrix, all with positive sign $\chi=1$. This implements the desired initial condition for \eq{bloch}.  The density matrix is then evolved in discrete steps of $\Delta \beta$, with $\beta / \Delta\beta$ steps taken. At each step, every psip $p$ (living on site $\left|b_p\right>\left<a_p\right|$) performs four operations:
\begin{enumerate}
	\item The psip may spawn a new psip on another site in the same column, $\left|c\right>\left<a_p\right|$ where $c \ne b_p$, with probability $\frac 1 2 \left|\left<c\right|H\left|b_p\right>\right|\Delta\beta$.
	\item Similarly, the psip may spawn a new psip onto another site in the same row, $\left|b_p\right>\left<c\right|$ where $c \ne a_p$, with probability $\frac 1 2 \left|\left<a_p\right|H\left|c\right>\right|\Delta\beta$.
	\item If $\left<a_p\right|H\left|a_p\right> + \left<b_p\right|H\left|b_p\right> > 0$, then the psip is removed from the simulation with probability $\frac 1 2 \left|\left<a_p\right|H\left|a_p\right> + \left<b_p\right|H\left|b_p\right>\right| \Delta \beta$.
	\item Alternatively, when $\left<a_p\right|H\left|a_p\right> + \left<b_p\right|H\left|b_p\right> < 0$, the psip is cloned, producing another psip on the same site. This occurs with probability $\frac 1 2 \left|\left<a_p\right|H\left|a_p\right> + \left<b_p\right|H\left|b_p\right>\right| \Delta \beta$.
\end{enumerate}

When the $\beta / \Delta\beta$ executions of these four steps have completed, the resulting collection of psips gives an approximation to $\rho(\beta)$ via \eq{rho}.

With the approximate density matrix $\tilde\rho$ determined, time-dependent expectation values are evaluated on a quantum processor.
A time-dependent expectation value is given by
\begin{equation}
	\left<\mathcal O(t)\right> = \Tr \mathcal O e^{-i H_1 t} \rho e^{i H_1 t}\text,
\end{equation}
where $H_1$, the Hamiltonian used for time evolution, is distinct from the $H_0$ Hamiltonian used to define the density matrix.
Substituting the hermitized approximate density matrix $\rho \rightarrow \frac 1 2 \left(\tilde \rho + \tilde\rho^\dagger\right)$, we see that the expectation value may be approximated by a sum over psips:
\begin{align}\label{eq:expval}
	&\left<\mathcal O(t)\right> \approx \frac 1 {\Tr \tilde\rho}\times\nonumber\\&
	\sum_p \Tr \bigg(\frac{1}{2} \mathcal O e^{-i H_1 t}
	\bigg[
		\chi_p \left|b_p\right>\left<a_p\right|+
		\bar\chi_p \left|a_p\right>\left<b_p\right|
		\bigg] e^{i H_1 t}\bigg)
\end{align}
From Eq.~\ref{eq:expval} it can be seen that the decomposition of the density matrix into psips allows  one to time-evolve each psip independently as a pure state, avoiding the difficulty of constructing a mixed state on a quantum processor.

Psips for which $a_p = b_p$ are termed `diagonal'. Expectation values $\left<a_p\right|\mathcal O(t)\left|a_p\right>$ of diagonal psips may be evaluated straightforwardly on a quantum computer because they can be represented easily as a pure state.  In contrast, non-diagonal psips must be diagonalized before evaluation on a quantum processor. For real charges $\chi_p$, a hermitized psip is diagonal in the basis $\left|u_p\right> = \left|a_p\right> + \left|b_p\right>$; $\left|v_p\right> = \left|a_p\right> - \left|b_p\right>$. Working in this basis (a different basis for each psip), the contribution to $\langle \mathcal{O}(t)\rangle$ of the non-diagonal psips becomes
\begin{align} \label{eq:expectation}
	&\left<\mathcal O(t)\right> \approx\nonumber\\&
	\frac 1 {\Tr \tilde\rho}
	\sum_p \left[
		\left<u\right| e^{i H_1 t} \mathcal O e^{-i H_1 t}\left|u\right>
		-\left<v\right| e^{i H_1 t} \mathcal O e^{-i H_1 t}\left|v\right>
	\right]\text.
\end{align}
In this form, the expectation value is a sum of quantities each amenable to computation with a quantum computer. For a given set of psips specifying $\tilde\rho$, a separate instance of a general program is run on the quantum processor for each psip. Each program contains the same code for time-evolution and measurement, but a different sequence of operations for preparing the pure states. For non-diagonal psips, two programs must be executed, one for $\left|u_p\right>$ and one for $\left|v_p\right>$, while the diagonal psips require only one. Each program has the following steps:
\begin{enumerate}
	\item Prepare the state $\left|u_p\right>$ (or $\left|v_p\right>$);
	\item Time-evolve with $H_1$ for a fixed time $t$ via trotterization;
	\item Measure $\mathcal O$, and any other observables of interest simultaneously.
\end{enumerate}

For nearly all Hamiltonians of physical interest, the diagonal basis of the Hamiltonian is not efficiently accessible, and the time-evolution operator $e^{i H_1 t}$ must be approximated by trotterization. This is accomplished by decomposing the Hamiltonian into terms easily diagonalized: $H_1 = H_x + H_z$. The time-evolution operator is then $e^{i H_1 t} = \left(e^{i H_x \Delta t} e^{i H_z \Delta t}\right)^{t/(\Delta t)} + O(\Delta t)$. In the case of \eq{hamiltonian}, we trotterize with $H_x = -\mu_x \sum_i \sigma^{(i)}_x$ and $H_z = -J_z \sum_{<ij>} \sigma^{(i)}_z \sigma^{(j)}_z - \mu_z \sum_i \sigma^{(i)}_z$.

In this paper, the observable of interest (transverse magnetization) may be measured by changing basis from the $Z-$ to the $X-$basis (a rotation of each qubit), and measuring all qubits simultaneously.



Once each psip has been evaluated by the quantum processor, the results are summed together (on the classical computer) via \eq{expectation} to calculate the expectation value of the thermal state.

The efficiency of this algorithm is strongly influenced by the fact that the approximate density matrix $\tilde\rho$ may be extremely sparse, where the exact density matrix $\rho$ is not. For an $N$-site system, the density matrix $\rho$ has at least $2^N$ non-zero entries; we expect sufficiently accurate expectation values to be obtainable with a population of psips which scales only polynomially with $N$. Each psip corresponds to one or two calculations on the quantum computer; thus, the number of calculations required on the quantum computer is expected to be polynomial in $N$. 

\section{The 1D Heisenberg Chain}\label{sec:model}
As a demonstration of the algorithm, we simulate a 1D time-dependent Heisenberg spin chain with one coupling constant and two independent magnetic fields~\cite{PhysRev.177.889,PhysRevE.71.036102,PhysRevB.77.064426,PhysRevLett.93.076401,1742-5468-2004-04-P04005}. The general Hamiltonian for this class of system is
\begin{equation}\label{eq:hamiltonian}
	H(t) = -J_z(t)\sum_{\left<ij\right>} \sigma^{(i)}_z \sigma^{(j)}_z - \mu_x(t) \sum_i \sigma^{(i)}_x - \mu_z(t) \sum_i \sigma^{(i)}_z\text,
\end{equation}
where $J_z(t)$ is the coupling constant between the $z-$axis aligned spin component of nearest neighbors, and $\mu_x(t)$ and $\mu_z(t)$ denote time-dependent magnetic fields aligned with the $x-$ and $z-$axes, respectively.  We take the spin chain to have periodic boundary conditions. In this paper, we will work in units where the inverse temperature is $\beta = 1$, and restrict ourselves to a constant coupling $J_z(t)=1$ and longitudinal magnetic field $\mu_z(t)$ which is $0$ for the $N=5$ system and $1$ for the $N=1$. The transverse magnetic field is permitted to be time-dependent.

The time-dependent observable we measure is the average transverse magnetization, given by
\begin{equation}
\mt\equiv \frac 1 N \sum_i\sigma^{(i)}_x(t)\text.
\end{equation}
As discussed in the previous section, this quantity is easily measured on the quantum processor.

\section{Results}\label{sec:results}
For the purposes of this exploratory study, we compute $\mt$ for two cases: the $N=5$ spin chain on the Rigetti Forest QVM to empirically test the algorithm's correctness, and the single-spin case on the Rigetti 8Q-Agave quantum computer to study the sources of uncertainty arising in a physical quantum processor.

\begin{figure}
	\centering
	\includegraphics[width=\linewidth]{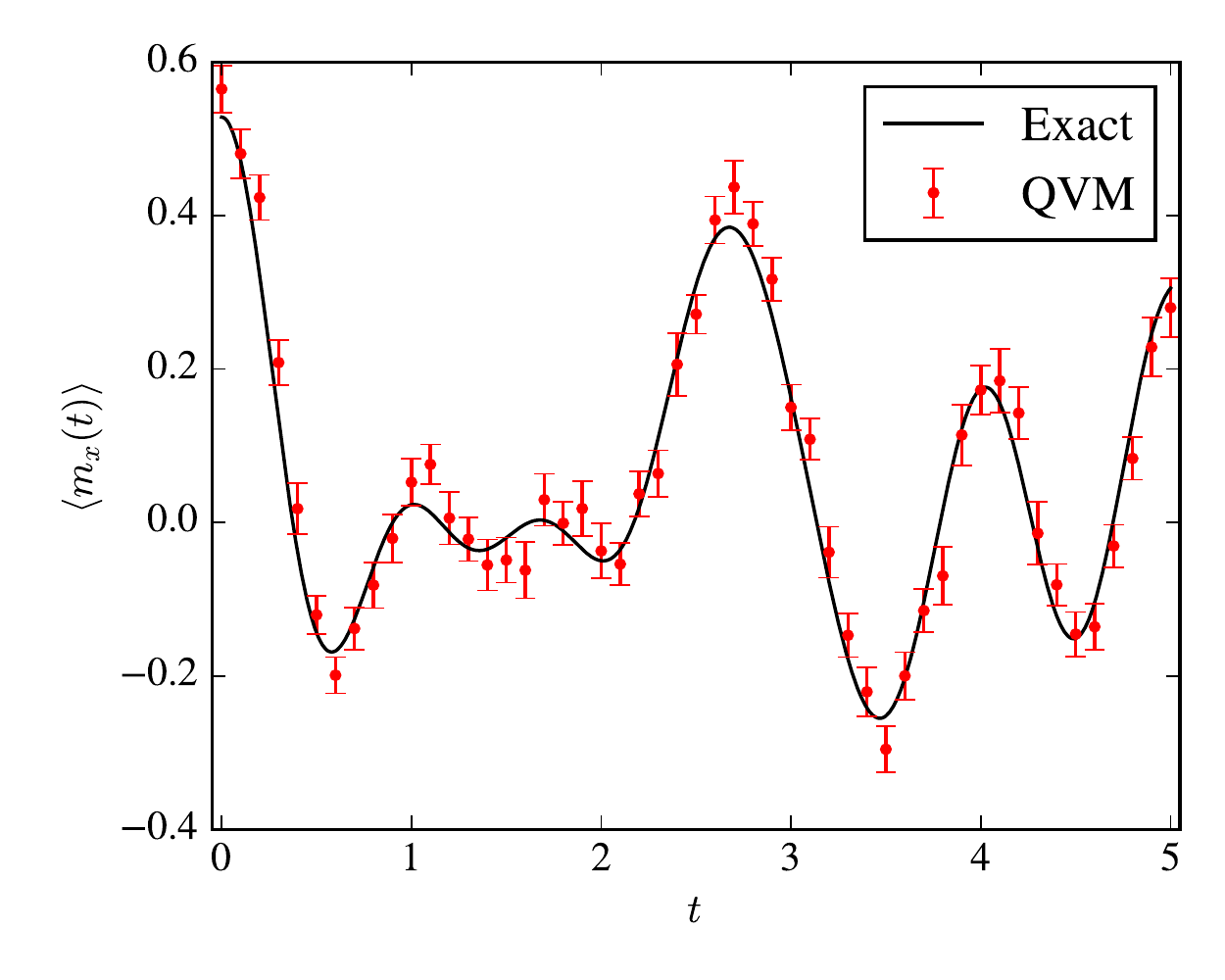}
	\caption{The transverse magnetization $\mt$ for a $N=5$ site spin chain with coupling $J_z=1$, and an initial $\mu_x(0)=1$ and $\beta=1$, which is evolved with $\mu_x(t> 0) = -1$. Results from the Forest QVM are shown by red circles and the exact result is denoted by the solid black line.\label{fig:qvm}}
\end{figure}

Without the additional sources of error inherent in a QPU, we are able to access larger systems on the QVM. We evolve the $N=5$ spin system with the Hamiltonian described by \eq{hamiltonian} with $\mu_x(t=0) = 1$ and $\mu_x(t>0) = -1$. The longitudinal magnetic field is $\mu_z = 0$.  For this calculation we use a trotterization time step of $\Delta t = 0.1$. The imaginary time step was $\Delta\beta = 0.04$ for evolving the psips with $5000$ initial psips. Shown in \fig{qvm} is $\mt$, in statistical agreement with the exact result.

When run on an ideal quantum processor, as simulated by Rigetti Forest, \eroq\ has two sources of uncertainty, both statistical: the approximation of $\rho$ by a finite number of psips, and the intrinsic measurement noise on the quantum processor.  These sources of error are easily accounted for with standard methods such as bootstrapping as we do in this work.  Note, though, that the errors are correlated since the same set of psips (i.e., the same approximation to the density matrix) is used for all values of $t$.

\begin{figure}
	\centering
	\includegraphics[width=\linewidth]{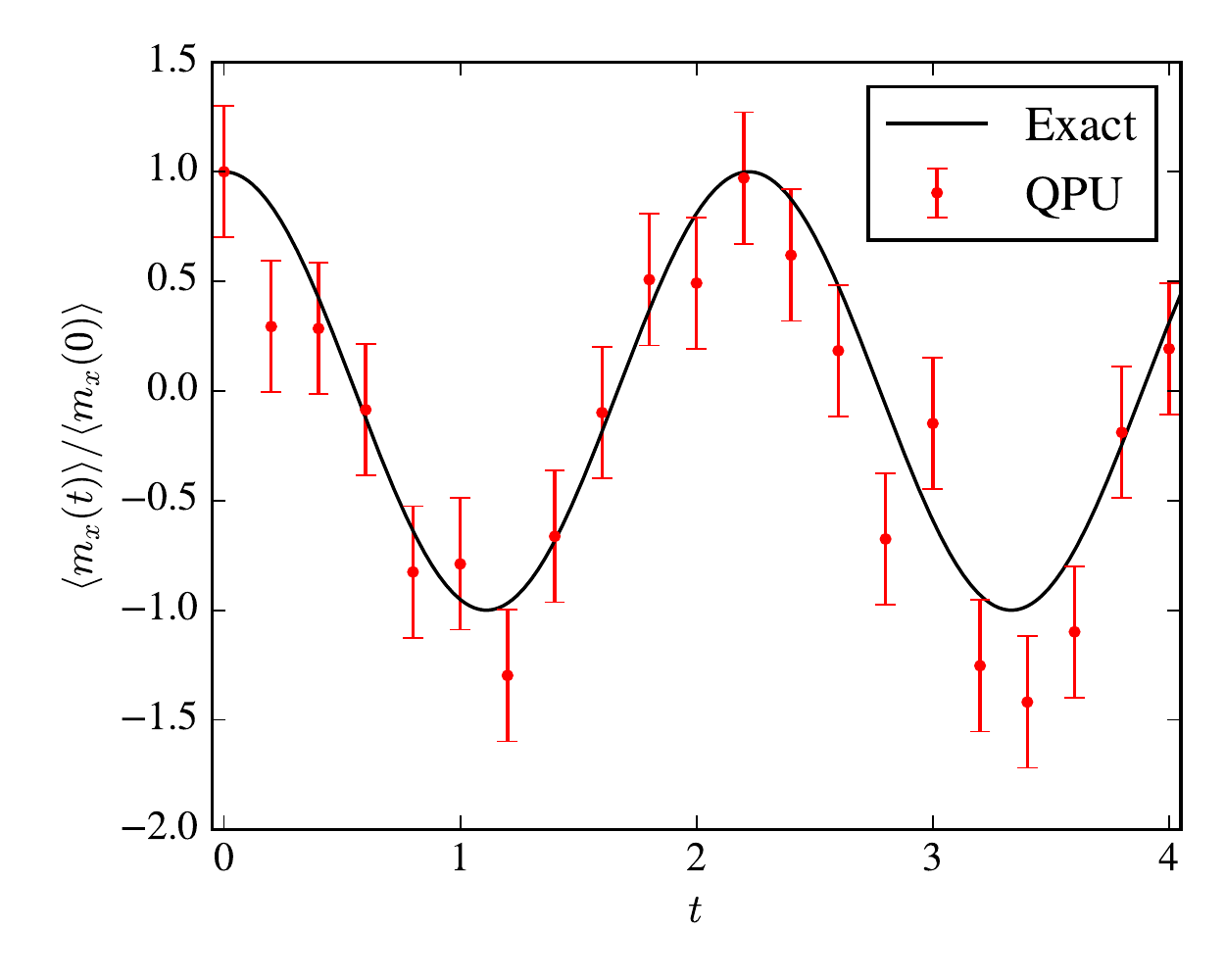}
	\caption{The rescaled (see text) transverse magnetization $\mt/\langle m_x(0)\rangle$ for a single spin, with initial $\mu_x(0)=\mu_z(0)=1$ and $\beta=1.0$, which is evolved with $\mu_x(t> 0) = -1$.  The results from Rigetti's 8Q-Agave QPU are shown in red circles while the exact result is denoted by the solid black line.\label{fig:qpu}}
\end{figure}

We use the 8-qubit quantum processor 8Q-Agave to simulate a single spin, thermalized in a transverse magnetic field $\mu_x(t=0)=1$, and time-evolved in a flipped magnetic field $\mu_x(t=0)=-1$.  The longitudinal magnetic field is taken to be constant: $\mu_z = 1$. For this calculation we use a trotterization time step of $\Delta t = 0.2$. The imaginary time step was $\Delta\beta = 0.04$, with $1000$ initial psips. The results of this execution of the algorithm are presented in \fig{qpu}, again in good agreement with the exact result.

The physical 8Q-Agave, unlike the simulated Forest, is not an ideal quantum processor, and has several additional sources of error that must be accounted for.  Most prominently, measurements have so-called {\em readout noise}. When measuring a qubit, there is some probability that the opposite state will be read instead.   If one assumes this readout noise is symmetric between the two states and independent of the gates used before a measurement is taken (empirically the case at our level of precision), this reduces the measured magnitude of $\mt$ by a constant factor, which can be corrected for by rescaling.  In \fig{qpu}, we rescale $\mt$ by $\langle m(0)\rangle$, which appears to sufficiently remove the effect of readout noise. 

Other sources of error, more difficult to correct for, are also present. For instance, when a parameterized gate (such as a 1-qubit phase gate) is requested with angle $\theta$, the actual gate implemented may have angle $\theta + \epsilon(\theta)$, producing a systematic bias in all results using that value of $\theta$. This and other unanticipated sources of systematic error may be accounted for by performing a calibration run with a simpler Hamiltonian (diagonal in the computational basis). For this work we use $H'_1 = - \mu_z \sigma_z$: the error bars estimated for \fig{qpu} are the quadrature average of the difference between the simulated results for $H'_1$ and the exact answer.

\section{Discussion and prospects}\label{sec:discussion}
In this work, we have presented \eroq\, a hybrid classical/quantum algorithm for simulating out-of-equilibrium dynamics of thermal quantum systems, applying it to a simple system on both a quantum virtual machine and a quantum processor. \eroq\ first computes an approximation of the density matrix upon a classical computer, evading the need to compute thermal physics or prepare a mixed state on a quantum computer.  The density matrix is then passed to a quantum processor to compute the time-evolution, thus avoiding the sign problem associated with real-time calculations on a classical computer.

Going forward, this algorithm could be applied to problems of greater physical interest. While the hadronization of the quark-gluon plasma or reheating in the early universe will require larger quantum processors than exist at present, the non-linear response of low-dimensional systems like spin chains and graphene as well as the response of a thermal neutron gas to neutrino scattering should be possible on near-future resources.  In order to do this, a better characterization of the errors present on today's physical quantum computers will be necessary --- a general concern for all quantum algorithms.

\begin{acknowledgments}
H.L. and S.L. are supported by the U.S. Department of Energy under Contract No.~DE-FG02-93ER-40762.  The authors would further like to thank Rigetti for their assistance and access to their resources, Forest and 8Q-Agave.
\end{acknowledgments}
\bibliographystyle{apsrev4-1}
\bibliography{wise}
\end{document}